\documentclass{Interspeech}
\usepackage{subcaption}

\makeatletter
\newcommand{\interspeechnote}{}

\let\old@maketitle\@maketitle
\def\@maketitle{%
  \old@maketitle
  \ifx\interspeechnote\@empty\else
    \vspace{-12mm}%
    \begin{center}
      \slshape\interspeechnote
    \end{center}%
    \vspace{2mm}%
  \fi
}
\makeatother

\interspeechcameraready

\title{Test-Time Training for Speech Enhancement}

\author[equalcontribution]{Avishkar}{Behera}
\author[equalcontribution]{Riya Ann}{Easow}
\author[]{Venkatesh}{Parvathala}
\author[]{K. Sri Rama}{Murty}

\affiliation[nocounter]{Speech Information Processing Lab, Department of Electrical Engineering}{Indian Institute of Technology Hyderabad}{India}
\email{\{avishkar112behera, riyaann76\}@gmail.com, ee22resch01005@iith.ac.in, ksrm@ee.iith.ac.in}


\usepackage{comment}
\usepackage{multirow}
\usepackage{float}
\usepackage{graphicx}
\begin{document}

\renewcommand{\interspeechnote}{Published in the Proceedings of Interspeech 2025 (DOI: 10.21437/Interspeech.2025-2725)}

\maketitle

\begin{abstract}
    This paper introduces a novel application of Test-Time Training (TTT) for Speech Enhancement, addressing the challenges posed by unpredictable noise conditions and domain shifts. This method combines a main speech enhancement task with a self-supervised auxiliary task in a Y-shaped architecture. The model dynamically adapts to new domains during inference time by optimizing the proposed self-supervised tasks like noise-augmented signal reconstruction or masked spectrogram prediction, bypassing the need for labeled data. We further introduce various TTT strategies offering a trade-off between adaptation and efficiency. Evaluations across synthetic and real-world datasets show consistent improvements across speech quality metrics, outperforming the baseline model. This work highlights the effectiveness of TTT in speech enhancement, providing insights for future research in adaptive and robust speech processing.
\end{abstract}

\section{Introduction}

Speech enhancement (SE) remains a critical challenge in speech signal processing, tasked with improving the perceptual quality and intelligibility of speech corrupted by noise. While deep neural networks (DNNs) have achieved state-of-the-art performance in denoising and source separation \cite{genhancer, yang2024SE, kong23c_interspeech}, their generalization to unseen noise environments remains limited. Supervised training on paired noisy-clean datasets yields strong performance under matched conditions, but models often degrade when encountering noise types absent from training data due to the domain mismatch between the noisy (simulated or real acoustic) conditions \cite{interNoise}.

Zhang et al. addressed partial domain adaptation (target domains with fewer classes than source domain) using importance-weighted adversarial networks \cite{zhang2018}. However, this approach does not mimic the real-world scenario where the speakers are independent and target domains are diverse. To solve this, Li et al. proposed a minimax method to transfer the model from a limited source domain to a rich target domain, strengthening generalization performance \cite{Li2022}. Learning Noise Adapters (LNAs) leveraged frozen pre-trained models with domain-specific adapters to incrementally handle new noise distributions while solving the catastrophic domain forgetting phenomenon \cite{LNA2024}. However, all the above methods require some information about the target noise or its distribution for training.

Recent advances in test-time adaptation (TTA) address the distribution shift issue through dynamic parameter adjustment during inference. This method involves adapting model weights without any additional training. For instance, Single-Utterance Test-Time Adaptation (SUTA) introduced by Lin et al. reduces ASR word error rates on out-of-domain speech by applying entropy minimization to adapt models per test utterance \cite{lin2022}. Another common TTA approach is to update the batch normalization layer statistics using a large number of test samples \cite{su2024BN, ttn_bn}. Furthermore, TTA frameworks like zero-shot knowledge distillation enable compact student models to adopt recurring noise/speaker patterns using pseudo-targets from larger teacher models, without clean reference signals \cite{kim2021test}.

Another paradigm employs weight prediction architectures in which auxiliary networks generate input-adaptive parameters. Dynamic Filter Networks (DFN) \cite{DFN2016} and WeightNet \cite{WeightNet2020} exemplify this approach by generating convolutional filters dynamically and predicting convolutional weights conditioned on input, respectively. Test-Time Training (TTT) emerges as a specialized form of this approach where we update weights through self-supervised objectives, helping in domain adaptation \cite{sun2020test}. The basic idea of TTT is to use a self-supervised task and a main task during training. During inference, the self-supervised task is used to fine-tune the model using test data, before making the main prediction. Section \ref{sec:TTT} discusses the TTT framework in further detail.

Dumpala et al. explored the application of TTT to speech classification tasks such as speaker identification and emotion detection \cite{TTTSpeech}. They identified key challenges with TTT, including sensitivity to optimization hyperparameters and scalability issues. To address these issues, they proposed a parameter-efficient fine-tuning (PEFT) algorithm that only considered the bias parameters for fine-tuning during TTT.

TTT addresses critical limitations of conventional methods through three key advantages. First, its domain-agnostic nature circumvents the need for explicit noise-type priors required by domain adversarial training (DAT), enabling adaptation to completely unseen noise distributions. Second, noise/speaker-informed weight prediction adapts models to the individual noise/speaker domains, giving superior performance even in challenging environments. Third, resource efficiency can be achieved through selective updates to bias terms or only a certain portion of the network, enabling real-time operation on edge devices.

Our goal is to develop a TTT-based SE framework that can generalize across diverse environments. We investigate different self-supervised tasks for TTT, analyzing their impact on noise suppression and overall speech enhancement quality. Additionally, we explore various TTT strategies, balancing the trade-offs between computational efficiency and the extent of performance improvement. To comprehensively evaluate the effectiveness of our approach, we conduct experiments on both synthetic and real datasets. By leveraging self-supervised learning, this approach has the potential to bridge the gap between training and test conditions, enabling more robust speech enhancement.

\section{Methodology}

\subsection{Test-Time Training (TTT) Framework}
\label{sec:TTT}

\begin{figure}
    \centering
    \begin{subfigure}[b]{0.595\linewidth}
        \centering
        \includegraphics[width=0.9\linewidth]{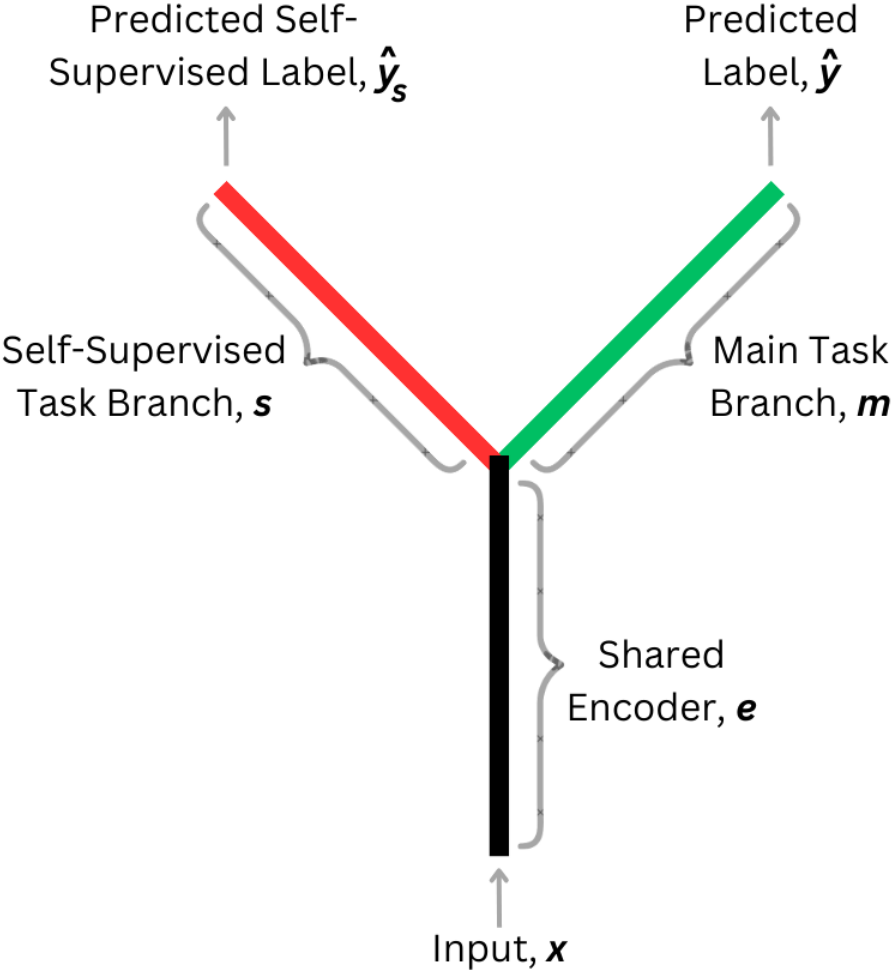}
        \caption{TTT architecture}
        \label{fig:TTT}
    \end{subfigure}
    \hfill
    \begin{subfigure}[b]{0.395\linewidth}
        \centering
        \includegraphics[width=0.9\linewidth]{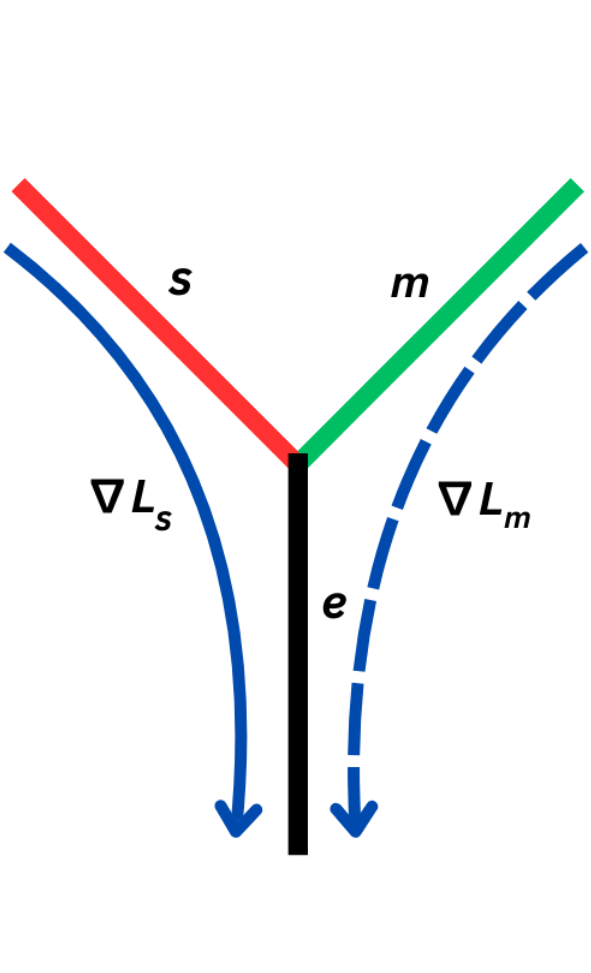}
        \caption{Gradient Flow Diagram}
        \label{fig:gradient}
    \end{subfigure}
    \caption{(a) Test-Time Training architecture and (b) Gradient backpropagation flow during training (both solid and dashed lines) and during testing (solid line only).}
\end{figure}

As proposed in Sun et al. \cite{sun2020test}, we use an architecture that consists of a shared encoder $e$ as the stem and two branches: a main task branch $m$ and a self-supervised auxiliary task branch $s$, thus giving the architecture a Y-shape as seen in Figure \ref{fig:TTT}. Let the parameters of the shared encoder, main task branch and self-supervised branch be $\theta_e$, $\theta_m$ and $\theta_s$ respectively. The main task loss ($\mathcal{L}_m$) is optimized over the parameters $\theta_e$ and $\theta_m$ and requires the input $x$ and label $y$. The self-supervised task loss ($\mathcal{L}_s$) does not use label $y$ but rather creates a pseudo-label $y_s$ from the unlabeled input $x$, and is optimized over the parameters $\theta_e$ and $\theta_s$.

During training, we jointly optimize over both the loss functions, $\mathcal{L}_s$ and $\mathcal{L}_m$ since the main task label $y$ is available to us as part of the training data ($x$, $y$). Hence, the training optimization problem becomes Eq.(\ref{eq:3}) as given in \cite{sun2020test}.
\begin{equation}
    \min_{\boldsymbol{\theta_e, \theta_m, \theta_s}} \mathbb{E} \left[ \mathcal{L}_m(x, y; \boldsymbol{\theta_e, \theta_m}) + \mathcal{L}_s(x; \boldsymbol{\theta_e, \theta_s}) \right]
    \label{eq:3}
\end{equation}

During testing, the main task label $y$ is not available but we can still optimize the self-supervised loss using the pseudo-label $y_s$. Hence, the testing optimization problem is the second term from Eq.(\ref{eq:3}). After optimization, we get the updated parameters, $\theta^*_e$ and $\theta^*_s$. Using the updated shared encoder ($\theta^*_e$) and the main task branch ($\theta_m$), we make the prediction i.e. $\hat{y}=\theta_{e^*+m}(x)$. In short, test-time training involves using the unlabeled input $x$ to update the model parameters via the self-supervised task, before making a prediction.

The gradient backpropagation flow is illustrated in Figure \ref{fig:gradient}. During training, gradients propagate from both the main branch $m$ and the self-supervised branch $s$ to the shared encoder $e$, updating the entire architecture. During testing, gradients flow only from the self-supervised branch $s$ to the shared encoder $e$, updating these components while keeping the main branch fixed.

We use the following four TTT strategies:

\begin{enumerate}
    \item \textbf{TTT-standalone:} This is the standard version of TTT where the model is initialized with parameters $\theta_e$ and $\theta_s$ obtained after training i.e. after optimizing Eq.(\ref{eq:3}). After updating the shared encoder using the test sample, we get $\theta^*_e$ which is used for prediction and then discarded. This is repeated for every sample $x_i$ in the test dataset.
    
    \item \textbf{TTT-online:} In the online version of TTT, the updated parameters $\theta^{*}_e$ and $\theta^{*}_s$ are not discarded, but instead used as initialization for the next test sample. That is, for a particular test sample $x_i$, the model parameters $\theta^{(i)}_e$ and $\theta^{(i)}_s$ are initialized using the updated parameters from the previous test sample i.e. $\theta^{*(i-1)}_e$ and $\theta^{*(i-1)}_s$.
    
    \item \textbf{TTT-online-batch:} This is similar to the TTT-online strategy but model parameters are updated using the current test sample as well as the previous four test samples. A batch size of 5 was empirically chosen as it yielded the highest performance improvement.
    
    \item \textbf{TTT-online-batch-bias:} Following the approach proposed in \cite{TTTSpeech}, only the bias parameters are updated during test time which reduces computational overhead while retaining adaptive capabilities. Here, we have implemented bias fine-tuning on top of the TTT-online-batch strategy.
\end{enumerate}

In our experiments, the main task is to denoise the noisy input signal by predicting a mask over the magnitude spectrogram. We have used two types of self-supervised tasks and analyzed their performances. The first self-supervised task is masked spectrogram prediction (\textbf{MSP}), where the model predicts the original spectrogram after random patches are zeroed out. This is similar to the MAE task proposed in \cite{zmolikova_MSP, gandelsman2022ttt, zhong_MSP}. The second type involves adding noise to the already noisy input signal, similar to the Noisy-target Training (\textbf{NyTT}) method in Fujimura et al. \cite{NyTT}. Hence, we get a noisier signal and the task is to denoise this augmented signal using the original noisy signal as reference.

\subsection{Model Architecture}
\label{sec:model_architecture}
In this paper, we will be using the real-time Neural Time-Varying Filtering (NTVF) network proposed by Venkatesh et al. in \cite{NTVF} for SE tasks. This network has shown substantial enhancement performance, while having significantly lower computational complexity. It takes the STFT log magnitude spectrogram $\log(|S[t,k]|)$ of the noisy signal as input and predicts a mask $\hat{W}[t,k]$. By multiplying the mask with the magnitude spectrogram, we get the enhanced magnitude spectrogram $|\hat{S}[t,k]|$. We reconstruct the enhanced time-domain signal by taking the inverse STFT using the noisy phase.

The typical NTVF network consists of a stack of $L$ identical blocks. We modify this network to the Y-shaped TTT framework depending on the self-supervised task. For the MSP self-supervised task, the shared encoder consists of $L$ = 4 blocks, the main branch consists of $L$ = 3 blocks followed by two dense layers with Tanh activation, and the self-supervised branch is identical to the main branch but uses ReLU instead.

For the NyTT self-supervised task, the shared encoder has $L$ = 6 blocks, the main task branch has a single block followed by two dense layers with Tanh activation, and the self-supervised task branch is identical to the main task branch. This is because both tasks are the same, i.e. to denoise the input signal. Hence, branching is done at a later stage with smaller main task and self-supervised task branches because we expect the shared encoder to extract similar features for both tasks.

\subsection{Loss Functions}
As suggested in \cite{NTVF}, we use a combination of spectral-related and perception-related losses to improve the enhancement quality. Namely, we use mask loss (mean absolute error between estimated and oracle TVF), SI-SDR loss and PESQ loss. Apart from these, we use the MSE (mean squared error) between the log magnitude spectrograms of the noisy input signal and the output of masked spectrogram prediction.

For the main task, we use a combination of mask loss, SI-SDR loss, and PESQ loss. For the MSP self-supervised task, we use MSE loss. For the NyTT self-supervised task, we use the same combination of losses as the main task since both tasks are identical.
\begin{align}
    \mathcal{L}_m &= \mathcal{L}_{Mask} + \mathcal{L}_{SI-SDR} + \mathcal{L}_{PESQ}\\
    \mathcal{L}_{s,MSP} &= \frac{1}{T.K} \sum^{T}_{t=1} \sum^{K}_{k=1} \left[\log|S[t,k]| - \log|\hat{S}[t,k]| \right]^2\\
    \mathcal{L}_{s,NyTT} &= \mathcal{L}_{Mask} + \mathcal{L}_{SI-SDR} + \mathcal{L}_{PESQ}
\end{align}

\section{Experiments and Results}

\subsection{Datasets}

\subsubsection{Training Data}
We use the DNS Challenge 1 dataset \cite{reddy2020interspeech2020deepnoise} to train the model. The real samples of this dataset are taken from the LibriVox corpus \cite{kearns2014librivox}. The noises are taken from Audioset \cite{gemmeke2017audio} and Freesound \cite{eduardo_fonseca_2017_1417159}. The dataset has about 150 audio classes including various noise types like music, wind, dog, etc. 500 hours of data was used to train the models.

\subsubsection{Test Data}
In our study, we have evaluated the different approaches on two types of data: synthetic data and real recordings.

For synthetic data, we assess the generalization ability of our model using the test set from the Valentini dataset \cite{valentinibotinhao16_interspeech}. The clean speech samples in this dataset are sourced from the Voice Bank Corpus \cite{veaux2013voice}, while the noisy speech signals are generated by adding various background noises from the DEMAND dataset \cite{thiemann2013diverse} at different signal-to-noise ratios (SNRs). These include real noises from settings such as bus, cafeteria, public square, living room and office. Overall, the 824 audio samples provided in the test set were used during evaluation. We observe an 8.01\% drop in PESQ when models are trained on the DNS dataset and evaluated on the Valentini test set, compared to models both trained and tested on Valentini. This highlights the domain shift between the two datasets.

While evaluating models on synthetic data provides insights, their performance in real-world scenarios may differ significantly. Therefore, we also evaluate our models on real recordings provided as part of the DNS Challenge 1 dataset, ensuring a more realistic assessment of our approach.

In order to test the extent of distribution shift caused by TTT, we re-evaluate our models on the DNS no-reverb test samples which consists of 300 audio samples.

\subsection{Experiments}

We experimented with three different self-supervised auxiliary tasks:  

\begin{enumerate}
    \item \textbf{MSP}: Masking the spectrogram of the noisy signal and predicting the spectrogram.
    \item \textbf{NyTT-gaussian}: Adding Gaussian noise to the noisy signal and predicting the original noisy signal.  
    \item \textbf{NyTT-real}: Adding real noises from river, park, field, domestic washing, and office hallway settings from the DEMAND dataset to the noisy signal and predicting the original noisy signal. The noise is added to the signal at an SNR randomly sampled between 0 and 15 dB.
\end{enumerate}

For each self-supervised task, we experimented with all the TTT strategies: TTT-standalone, TTT-online, TTT-online-batch and TTT-online-batch-bias. These experiments are compared with the NVTF model (\textbf{Baseline}) that is trained without any self-supervised auxiliary task and the NVTF model trained jointly with the main task and the self-supervised task (\textbf{Joint Training}).

\subsection{Experimental Configuration}
The model configuration follows the setup described in Section \ref{sec:model_architecture}. We use the AdamW optimizer \cite{loshchilovadamw} for training with an initial learning rate of 0.001. To adaptively adjust the learning rate based on performance, we employ a learning rate scheduler to reduce the learning rate when it plateaus. All models were trained for 100 epochs.

While evaluating the model, we empirically choose a learning rate of 1e-4 for the Valentini dataset and 1e-6 for the real recordings for the MSP and NyTT-real auxiliary tasks. For NyTT-gaussian, we have chosen a learning rate of 1e-6 for both the datasets. 

\subsection{Evaluation Metrics}

{\setlength{\abovecaptionskip}{7.5pt}%
\begin{table*}[ht!]
\centering
\setlength{\tabcolsep}{12pt}
\caption{Performance comparison across Valentini dataset and real audios with various TTT strategies.}
\begin{tabular}{llcccccc}
\toprule
\textbf{Experiment Type} & \textbf{Method} & \multicolumn{3}{c}{\textbf{Valentini}} & \multicolumn{3}{c}{\textbf{Real Audios}} \\
\cmidrule(lr){3-5} \cmidrule(lr){6-8}
& & \textbf{PESQ} & \textbf{STOI} & \textbf{SSNR} & \textbf{DSIG} & \textbf{DBAK} & \textbf{DOVRL} \\
\midrule
\multirow{1}{*}{Noisy} & - & 1.971 & 92.099 & 1.680 & 3.053 & 2.509 & 2.255 \\
\midrule
\multirow{1}{*}{Baseline} & NVTF & 2.961 & 92.972 & 8.684 & 3.290 & 3.950 & 2.984 \\
\midrule
\multirow{5}{*}{MSP} 
& Joint Training & 2.956 & 92.985 & 8.811 & 3.300 & 3.926 & 2.980 \\
& TTT-standalone & 2.958 & 93.004 & 8.819 & 3.300 & 3.926 & 2.980 \\
& TTT-online & 3.004 & 93.324 & 8.831 & 3.299 & 3.928 & 2.979 \\
& TTT-online-batch & 3.034 & 92.213 & 8.948 &  3.299 & 3.928 & 2.979 \\
& TTT-online-batch-bias & 3.015 & 92.049 & 8.969 &  3.300 & 3.927 & 2.980 \\
\midrule
\multirow{5}{*}{NyTT-gaussian} 
& Joint Training & 2.983 & 93.261 & 9.248 & 3.317 & \textbf{3.982} & 3.018 \\
& TTT-standalone & 2.983 & 93.262 & 9.248 & 3.317 & \textbf{3.982} & 3.018 \\
& TTT-online & 2.948 & 93.337 & 9.306 & 3.318 & \textbf{3.982} & \textbf{3.019} \\
& TTT-online-batch & 2.971 & 92.397 & \textbf{9.518} & 3.317 & \textbf{3.982} & \textbf{3.019} \\
& TTT-online-batch-bias & 2.999 & 92.276 & 9.457 & 3.317 & 3.981 & \textbf{3.019} \\

\midrule
\multirow{5}{*}{NyTT-real} 
& Joint Training & 2.981 & \textbf{93.493} & 9.068 & 3.319 & 3.950 & 3.007 \\
& TTT-standalone & 2.982 & 93.484 & 9.066 & 3.320 & 3.950 & 3.007 \\
& TTT-online & 3.081 & 93.323 & 8.469 & \textbf{3.324} & 3.954 & 3.012 \\
& TTT-online-batch & \textbf{3.145} & 92.643 & 8.932 & \textbf{3.324} & 3.954 & 3.013 \\
& TTT-online-batch-bias  & 3.088 & 92.515 & 9.038 & 3.321 & 3.951 & 3.008 \\
\bottomrule
\end{tabular}
\label{tab:valentini_real_audio_results}
\end{table*}
}

We evaluate the model using the following measures: Perceptual Evaluation of Speech Quality (PESQ) \cite{rix_pesq}, Short-Time Objective Intelligibility (STOI) \cite{taal2010short}, and Segmental Signal-to-Noise Ratio (SSNR) \cite{hansen1998effective}, all of which require a clean reference signal for comparison.

We use DSIG (speech quality assessment), DBAK (background noise suppression evaluation), and DOVRL (overall audio quality assessment) from DNSMOS \cite{reddy2022dnsmos}, a non-intrusive neural network based metric for evaluating speech quality. These scores do not require a reference signal, making them particularly useful for real-world recordings where a clean reference may not be available. Each score is rated on a scale from 1 to 5.

\subsection{Results}
The performance of the proposed methods across various self-supervised tasks are shown in Table \ref{tab:valentini_real_audio_results}.

\begin{enumerate}
    \item \textbf{Valentini dataset}: Joint training with auxiliary tasks (except MSP) provides slight but consistent gains across all metrics over the baseline, suggesting that integrating an auxiliary self-supervised objective benefits the main task in generalizing to another domain.
        
    In the MSP task, TTT and TTT-online lead to an improvement of all metrics. For NyTT-gaussian, these strategies cause a slight drop in PESQ but improve SSNR. Conversely, for NyTT-real, they slightly reduce SSNR but improve PESQ.
    
     Across all auxiliary tasks, TTT-online-batch enhances both the PESQ and SSNR compared to TTT-online but introduces a negligible drop in STOI. Further, our experiments indicate that using a batch of the previous four samples yields better performance than increasing either the number of gradient steps per sample or the learning rate, as it allows the model to capture the underlying data distribution rather than overfitting to the noise characteristics of a single speech sample. TTT-online-batch-bias results in a minor PESQ drop but compensates with slight SSNR improvements, indicating that restricting adaptation to bias parameters preserves stability while offering computation benefits.

    \item \textbf{Real audios}: On the real dataset, the MSP task slightly increases the DSIG but fails to suppress the background noise (captured by the low DBAK). This results in a lower DOVRL score. This suggests that masked spectrogram prediction may not provide meaningful adaptation cues in real-world conditions. In contrast, the NyTT-gaussian experiment significantly enhances DBAK, leading to the highest DOVRL score among all methods. This indicates that adding Gaussian noise effectively improves background noise suppression. On the other hand, the NyTT-real experiment results in noticeable improvements in DSIG, indicating better preservation of speech content. This improvement further translates to a higher DOVRL score compared to the baseline.

\end{enumerate}

\subsection{Extent of Distribution Shift after TTT}
After evaluating the model on the Valentini dataset, we re-evaluated it on the DNS test set without any TTT strategy to assess the extent of distributional shift from the original training data for the various TTT strategies. We selected the best-performing model on Valentini, the NyTT-real model, for this experiment. 

{\setlength{\abovecaptionskip}{7.5pt}%
\begin{table}[ht!]
\centering
\caption{Performance of NyTT-real (without weight updation) on the DNS test set after TTT on the Valentini test set.}
\resizebox{\columnwidth}{!}{%
\begin{tabular}{lccc}
\toprule
\textbf{Method} & \textbf{PESQ} & \textbf{STOI} & \textbf{SSNR} \\
\midrule
Noisy & 1.582 & 91.522 & 5.965 \\
\midrule
NyTT-real Joint Training & 3.126 & 96.475 & 10.736 \\
NyTT-real TTT-standalone & 3.131 & 96.481 & 10.611 \\
NyTT-real TTT-online & 3.094 & 96.263 & 10.145 \\
NyTT-real TTT-online-batch & 3.065 & 96.305 & 10.427 \\
NyTT-real TTT-online-batch-bias & 3.093 & 96.421 & 10.623 \\
\bottomrule
\end{tabular}
}
\label{tab:nytt_real_ttt_results}
\end{table}
}

As seen in Table \ref{tab:nytt_real_ttt_results}, performance on the DNS test set degrades after TTT on the Valentini test set, indicating a shift in distribution. The smallest drop is observed in TTT-standalone and the largest drop is observed in TTT-online-batch. This reinforces that a single adaptation step minimizes deviation from the original model while performing TTT with more samples leads to greater distributional shift. However, TTT-online-batch-bias mitigates this drop slightly, reinforcing that restricting updates to bias parameters helps preserve generalization while still leveraging test-time adaptation. These results show that our model does not suffer from the catastrophic domain forgetting phenomenon, unlike other domain adaptation models.

\section{Conclusion and Future Work}

The integration of test-time training into speech enhancement represents a paradigm shift from static models to adaptive systems that continuously refine parameters against real-world variability. Our experiments show that TTT improves speech quality across both synthetic (Valentini) and real-world (DNS test) datasets. Among the self-supervised tasks, NyTT-real excels at preserving speech quality, while NyTT-gaussian is more effective at suppressing background noise. This highlights a trade-off between speech quality and noise suppression, suggesting that task selection should be application-dependent. For adaptation strategies, TTT-online-batch provides significant gains reducing the PESQ drop observed when training on DNS and testing on Valentini from 8.01\% to 2.09\%, effectively mitigating the impact of domain shift. On the other hand, TTT-online-batch-bias offers a computationally efficient alternative. We have also shown that the model adapted to a new domain using TTT, still retains significant performance on the source domain. This framework can be integrated with any SOTA speech enhancement model, and further extended to tasks like personalized speech enhancement by training on a single speaker and adapting to new speakers at test time.

\bibliographystyle{IEEEtran}
\bibliography{mybib}

\begin{thebibliography}{10}
\providecommand{\url}[1]{#1}
\csname url@samestyle\endcsname
\providecommand{\newblock}{\relax}
\providecommand{\bibinfo}[2]{#2}
\providecommand{\BIBentrySTDinterwordspacing}{\spaceskip=0pt\relax}
\providecommand{\BIBentryALTinterwordstretchfactor}{4}
\providecommand{\BIBentryALTinterwordspacing}{\spaceskip=\fontdimen2\font plus
\BIBentryALTinterwordstretchfactor\fontdimen3\font minus \fontdimen4\font\relax}
\providecommand{\BIBforeignlanguage}[2]{{%
\expandafter\ifx\csname l@#1\endcsname\relax
\typeout{** WARNING: IEEEtran.bst: No hyphenation pattern has been}%
\typeout{** loaded for the language `#1'. Using the pattern for}%
\typeout{** the default language instead.}%
\else
\language=\csname l@#1\endcsname
\fi
#2}}
\providecommand{\BIBdecl}{\relax}
\BIBdecl

\bibitem{genhancer}
H.~Yang, J.~Su, M.~Kim, and Z.~Jin, ``Genhancer: High-fidelity speech enhancement via generative modeling on discrete codec tokens,'' in \emph{Interspeech 2024}, 2024, pp. 1170--1174.

\bibitem{yang2024SE}
Y.~Yang, N.~Trigoni, and A.~Markham, ``Pre-training feature guided diffusion model for speech enhancement,'' in \emph{Interspeech 2024}, 2024, pp. 1185--1189.

\bibitem{kong23c_interspeech}
Z.~Kong, W.~Ping, A.~Dantrey, and B.~Catanzaro, ``{CleanUNet 2}: A hybrid speech denoising model on waveform and spectrogram,'' in \emph{Interspeech 2023}, 2023, pp. 790--794.

\bibitem{interNoise}
J.~CUI and S.~BLEECK, ``From simulation to reality: tackling data mismatches in speech enhancement with unsupervised pre-training,'' \emph{INTER-NOISE and NOISE-CON Congress and Conference Proceedings}, vol. 270, no.~11, pp. 206--215, 2024.

\bibitem{zhang2018}
J.~Zhang, Z.~Ding, W.~Li, and P.~Ogunbona, ``Importance weighted adversarial nets for partial domain adaptation,'' in \emph{Proceedings of the IEEE conference on computer vision and pattern recognition}, 2018, pp. 8156--8164.

\bibitem{Li2022}
Y.~Li, Y.~Sun, K.~Horoshenkov, and S.~M. Naqvi, ``{Domain Adaptation and Autoencoder-Based Unsupervised Speech Enhancement},'' \emph{IEEE Transactions on Artificial Intelligence}, vol.~3, no.~1, pp. 43--52, 2022.

\bibitem{LNA2024}
Z.~Yang, X.~Song, J.~Chen, C.~Richard, and I.~Cohen, ``{Learning Noise Adapters for Incremental Speech Enhancement},'' \emph{IEEE Signal Processing Letters}, vol.~31, pp. 2915--2919, 2024.

\bibitem{lin2022}
G.-T. Lin, S.-W. Li, and H.~yi~Lee, ``{Listen, Adapt, Better WER}: Source-free single-utterance test-time adaptation for automatic speech recognition,'' in \emph{Interspeech 2022}, 2022, pp. 2198--2202.

\bibitem{su2024BN}
Z.~Su, J.~Guo, K.~Yao, X.~Yang, Q.~Wang, and K.~Huang, ``Unraveling batch normalization for realistic test-time adaptation,'' in \emph{Proceedings of the AAAI Conference on Artificial Intelligence}, vol.~38, no.~13, 2024, pp. 15\,136--15\,144.

\bibitem{ttn_bn}
H.~Lim, B.~Kim, J.~Choo, and S.~Choi, ``{TTN}: A domain-shift aware batch normalization in test-time adaptation,'' in \emph{Proceedings of the 11th International Conference on Learning Representations (ICLR)}, 2023.

\bibitem{kim2021test}
S.~Kim and M.~Kim, ``Test-time adaptation toward personalized speech enhancement: Zero-shot learning with knowledge distillation,'' in \emph{2021 IEEE Workshop on Applications of Signal Processing to Audio and Acoustics (WASPAA)}.\hskip 1em plus 0.5em minus 0.4em\relax IEEE, 2021, pp. 176--180.

\bibitem{DFN2016}
X.~Jia, B.~De~Brabandere, T.~Tuytelaars, and L.~V. Gool, ``Dynamic filter networks,'' \emph{Advances in neural information processing systems}, vol.~29, 2016.

\bibitem{WeightNet2020}
N.~Ma, X.~Zhang, J.~Huang, and J.~Sun, ``{WeightNet}: Revisiting the design space of weight networks,'' in \emph{Computer Vision -- ECCV 2020}, A.~Vedaldi, H.~Bischof, T.~Brox, and J.-M. Frahm, Eds.\hskip 1em plus 0.5em minus 0.4em\relax Cham: Springer International Publishing, 2020, pp. 776--792.

\bibitem{sun2020test}
Y.~Sun, X.~Wang, Z.~Liu, J.~Miller, A.~A. Efros, and M.~Hardt, ``Test-time training with self-supervision for generalization under distribution shifts,'' in \emph{Proceedings of the 37th International Conference on Machine Learning}, ser. ICML'20.\hskip 1em plus 0.5em minus 0.4em\relax JMLR.org, 2020.

\bibitem{TTTSpeech}
S.~H. Dumpala, C.~Sastry, and S.~Oore, ``Test-time training for speech,'' \emph{arXiv preprint arXiv:2309.10930}, 2023.

\bibitem{zmolikova_MSP}
K.~Zmolikova, M.~S. Pedersen, and J.~Jensen, ``Masked spectrogram prediction for unsupervised domain adaptation in speech enhancement,'' \emph{IEEE Open Journal of Signal Processing}, vol.~5, pp. 274--283, 2024.

\bibitem{gandelsman2022ttt}
Y.~Gandelsman, Y.~Sun, X.~Chen, and A.~A. Efros, ``Test-time training with masked autoencoders,'' in \emph{Advances in Neural Information Processing Systems}, A.~H. Oh, A.~Agarwal, D.~Belgrave, and K.~Cho, Eds., 2022.

\bibitem{zhong_MSP}
Z.~Zhong, H.~Shi, M.~Hirano, K.~Shimada, K.~Tateishi, T.~Shibuya, S.~Takahashi, and Y.~Mitsufuji, ``Extending audio masked autoencoders toward audio restoration,'' in \emph{2023 IEEE Workshop on Applications of Signal Processing to Audio and Acoustics (WASPAA)}, 2023, pp. 1--5.

\bibitem{NyTT}
T.~Fujimura and T.~Toda, ``Analysis of noisy-target training for dnn-based speech enhancement,'' in \emph{ICASSP 2023 - 2023 IEEE International Conference on Acoustics, Speech and Signal Processing (ICASSP)}, 2023, pp. 1--5.

\bibitem{NTVF}
V.~Parvathala and S.~R.~M. Kodukula, ``Light-weight causal speech enhancement using time-varying multi-resolution filtering,'' in \emph{2024 National Conference on Communications (NCC)}, 2024, pp. 1--6.

\bibitem{reddy2020interspeech2020deepnoise}
C.~K. Reddy, V.~Gopal, R.~Cutler, E.~Beyrami, R.~Cheng, H.~Dubey, S.~Matusevych, R.~Aichner, A.~Aazami, S.~Braun, P.~Rana, S.~Srinivasan, and J.~Gehrke, ``The {INTERSPEECH 2020} deep noise suppression challenge: Datasets, subjective testing framework, and challenge results,'' in \emph{Interspeech 2020}, 2020, pp. 2492--2496.

\bibitem{kearns2014librivox}
J.~Kearns, ``Librivox: Free public domain audiobooks,'' \emph{Reference Reviews}, vol.~28, no.~1, pp. 7--8, 2014.

\bibitem{gemmeke2017audio}
J.~F. Gemmeke, D.~P. Ellis, D.~Freedman, A.~Jansen, W.~Lawrence, R.~C. Moore, M.~Plakal, and M.~Ritter, ``Audio set: An ontology and human-labeled dataset for audio events,'' in \emph{2017 IEEE international conference on acoustics, speech and signal processing (ICASSP)}.\hskip 1em plus 0.5em minus 0.4em\relax IEEE, 2017, pp. 776--780.

\bibitem{eduardo_fonseca_2017_1417159}
\BIBentryALTinterwordspacing
E.~Fonseca, J.~Pons, X.~Favory, F.~Font, D.~Bogdanov, A.~Ferraro, S.~Oramas, A.~Porter, and X.~Serra, ``Freesound datasets: A platform for the creation of open audio datasets.'' in \emph{Proceedings of the 18th International Society for Music Information Retrieval Conference}.\hskip 1em plus 0.5em minus 0.4em\relax ISMIR, Oct. 2017, pp. 486--493. [Online]. Available: \url{https://doi.org/10.5281/zenodo.1417159}
\BIBentrySTDinterwordspacing

\bibitem{valentinibotinhao16_interspeech}
C.~Valentini-Botinhao, X.~Wang, S.~Takaki, and J.~Yamagishi, ``{Speech Enhancement for a Noise-Robust Text-to-Speech Synthesis System Using Deep Recurrent Neural Networks},'' in \emph{Interspeech 2016}, 2016, pp. 352--356.

\bibitem{veaux2013voice}
C.~Veaux, J.~Yamagishi, and S.~King, ``The voice bank corpus: Design, collection and data analysis of a large regional accent speech database,'' in \emph{2013 International Conference Oriental COCOSDA held jointly with 2013 Conference on Asian Spoken Language Research and Evaluation (O-COCOSDA/CASLRE)}, 2013, pp. 1--4.

\bibitem{thiemann2013diverse}
J.~Thiemann, N.~Ito, and E.~Vincent, ``The diverse environments multi-channel acoustic noise database (demand): A database of multichannel environmental noise recordings,'' in \emph{Proceedings of Meetings on Acoustics}, vol.~19, no.~1.\hskip 1em plus 0.5em minus 0.4em\relax AIP Publishing, 2013.

\bibitem{loshchilovadamw}
I.~Loshchilov and F.~Hutter, ``Decoupled weight decay regularization,'' in \emph{International Conference on Learning Representations}, 2019.

\bibitem{rix_pesq}
A.~Rix, J.~Beerends, M.~Hollier, and A.~Hekstra, ``{Perceptual evaluation of speech quality (PESQ)}-a new method for speech quality assessment of telephone networks and codecs,'' in \emph{2001 IEEE International Conference on Acoustics, Speech, and Signal Processing. Proceedings (Cat. No.01CH37221)}, vol.~2, 2001, pp. 749--752 vol.2.

\bibitem{taal2010short}
C.~H. Taal, R.~C. Hendriks, R.~Heusdens, and J.~Jensen, ``A short-time objective intelligibility measure for time-frequency weighted noisy speech,'' in \emph{2010 IEEE international conference on acoustics, speech and signal processing}.\hskip 1em plus 0.5em minus 0.4em\relax IEEE, 2010, pp. 4214--4217.

\bibitem{hansen1998effective}
J.~H.~L. Hansen and B.~L. Pellom, ``An effective quality evaluation protocol for speech enhancement algorithms,'' in \emph{5th International Conference on Spoken Language Processing (ICSLP 1998)}, 1998, p. paper 0917.

\bibitem{reddy2022dnsmos}
C.~K. Reddy, V.~Gopal, and R.~Cutler, ``{DNSMOS P. 835}: A non-intrusive perceptual objective speech quality metric to evaluate noise suppressors,'' in \emph{ICASSP 2022-2022 IEEE International Conference on Acoustics, Speech and Signal Processing (ICASSP)}.\hskip 1em plus 0.5em minus 0.4em\relax IEEE, 2022, pp. 886--890.

\end{thebibliography}

\end{document}